\documentclass[aps, prd, groupedaddress, preprint, tightenlines,  eqsecnum, showpacs, nofootinbib]{revtex4}

\usepackage{natbib}

\usepackage{epsfig}
\usepackage{graphicx}

\usepackage{axodraw2, float, slashed, graphicx, amssymb, amsmath}
\usepackage[usenames, dvipsnames]{xcolor}

\def\Li{\mathop{\hbox{\rm Li}}\nolimits}

\def\ksl{\slashed{k}}

\def\Tr{{\rm Tr}}

\def\Fcc{P^{(2)}}

\def\spa#1.#2{\left\langle#1\,#2\right\rangle}
\def\spb#1.#2{\left[#1\,#2\right]}
\def\la{\langle}
\def\ra{\rangle}

\DeclareMathOperator{\tr}{\mathrm{Tr}}

\def\PT{C_{PT}}
\def\CY{Cy}

\def\be{\begin{equation}}
\def\ee{\end{equation}}

\begin{document}

\hfill\today

\title{ An $n$-point QCD two-loop amplitude}

\author{David~C.~Dunbar, Warren~B.~Perkins and  Joseph~M.W.~Strong}

\affiliation{
College of Science, \\
Swansea University, \\
Swansea, SA2 8PP, UK\\
\today
}

\begin{abstract}
We present an explicit expression for a particular $n$-gluon two loop scattering partial amplitude.
Specifically we present an analytic form for the single trace $N_c$ independent colour partial amplitude for the case where all external gluons have positive helicity. 
\end{abstract}


\maketitle

\section{Introduction}

Computing scattering amplitudes is a key technology in producing theoretical predictions to test at colliders and other experiments. 
With increasing experimental data there is an insatiable demand for more and more 
accurate theoretical predictions~\cite{Bendavid:2018nar,Azzi:2019yne},
particularly for gauge theory amplitudes.
Amplitudes are also of more formal interest in that they exhibit the full symmetries of the theory. 
Unfortunately, these are not
easy to generate although great progress has been made in the last few years.

In a Yang-Mills gauge theory a $n$-gluon amplitude may be expanded in the gauge coupling constant,
\begin{equation}
{\cal A}_n = g^{n-2}  \sum_{\ell\geq 0} a^{\ell}{\cal A}_n^{(\ell)} 
\end{equation}
where $a=g^2e^{-\gamma_E \epsilon}/(4\pi)^{2-\epsilon}$.  In $SU(N_c)$ and $U(N_c)$ gauge theories a loop amplitude can be
further expanded in terms of color structures, $C^\lambda$,
\begin{equation}
{\cal A}_n^{(\ell)} = \sum_{\lambda} { A}_{n:\lambda}^{(\ell)} C^\lambda\,,
\end{equation}
separating the color and kinematics of the amplitude. The color structures $C^\lambda$ may be organised in terms of powers of $N_c$. 

There has been much progress in computing leading (tree, $\ell=0$) and ``next-to-leading order'', (one loop, $\ell=1$)
amplitudes. For ``next-to-next-to leading order''
progress has been considerable in theories with highly extended supersymmetry, both at the integrated~\cite{Caron-Huot:2019vjl} and integrand level~\cite{Bourjaily:2019gqu}. 
However for pure gauge theory progress
has  been restricted to amplitudes with a small number of external legs. Specifically  full results are only
available analyically for four gluons~\cite{Glover:2001af,Bern:2002tk}, and in~\cite{Ahmed:2019qtg} to all orders in dimensional regularisation.
For five external gluons
progress has focussed upon dividing the full amplitude into its different color and helicity partial amplitudes.
The first amplitude to be computed at five point was the leading in 
color part of the amplitude with all positive helicity external gluons (the all-plus amplitude) 
which was computed using $d$-dimensional  unitarity methods~\cite{Badger:2013gxa,Badger:2015lda}
and was subsequently presented in a very elegant and compact form~\cite{Gehrmann:2015bfy}. 
In~\cite{Dunbar:2016aux}, it was shown how four-dimensional unitarity techniques could be used to regenerate the five-point leading in color amplitude. The leading in color five-point amplitudes have been computed for the remaining helicities~\cite{Abreu:2019odu,Badger:2018enw}.  
Full color amplitudes are significantly more complicated requiring a larger class of master integrals incorporating 
non-planar integrals~\cite{Chicherin:2018old,Chawdhry:2018awn}. In~\cite{Badger:2019djh} the first full color five-point amplitude was presented in QCD -for the case of all-plus helicities.   
Beyond five-point, only the leading in color all-plus amplitudes for six- and seven-points are known~\cite{Dunbar:2016gjb,Dunbar:2017nfy}. 
 
In this article, we will present a conjecture for a very specific color partial two-loop amplitude which is valid for an arbitrary number of legs.  Again, it will be the case where all external gluons
have positive helicity: this being the most symmetric combination.   
The specific color structure is in many ways the most sub-leading term where there are no factors of $N_c$ and a single trace of the color matrices.  From a very different viewpoint,
this partial amplitude arises in open string theory from the non-planar  two loop orientable surface.   
Although it is very specific (and probably the least important
phenomenologically)  this hopefully will provide a useful multi-leg two loop expression from which to study the structure and properties of amplitudes.

\section{Color structures of Amplitudes} 

For completeness we review some aspects of tree and loop amplitudes which we will refer to later.
An $n$-point tree amplitude can be expanded in a color trace basis as
\begin{eqnarray}
{\cal A}_n^{(0)}(1,2,3,\cdots ,n)  &=& \sum_{S_n/Z_n}  \Tr[ T^{a_1} \cdots T^{a_n}] A_{n}^{(0)} (a_1,a_2,\cdots, a_n).
\end{eqnarray}
This separates the color and kinematic structures.
The partial amplitudes $A_{n}^{(0)} (a_1,a_2,\cdots, a_n)$ are cyclically symmetric but not fully crossing symmetric, 
they are however fully gauge invariant.  
The sum over permutations is over the $(n-1)!$
permutations of $(1,2,\cdots, n)$ up to this cyclic symmetry (this is not the only expansion, others exist~\cite{DelDuca:1999rs}
which  may be more efficient for some purposes).  This color decomposition is valid for both  $U(N_c)$ and $SU(N_c)$ gauge theories.  
If any of the external particles in the $U(N_c)$ case are $U(1)$ particles then the amplitude must vanish.  
This imposes {\it decoupling identities} amongst the partial amplitudes~\cite{Bern:1990ux}.  
For example setting  leg $1$ to be $U(1)$ and extracting the coefficient of 
$\Tr[T^2 T^3 \cdots T^n]$ implies that
\begin{equation}
A_{n}^{(0)}(1,2,3,\cdots , n ) +A_{n}^{(0)}(2,1,3,\cdots  , n )+\cdots A_{n}^{(0)}(2,\cdots, 1, n )=0.
\label{eq:decotree}
\end{equation}

The  one-loop $n$-point amplitude can be expanded as~\cite{Bern:1990ux}
\begin{eqnarray}
& & {\cal A}_n^{(1)}(1,2,3,\cdots ,n)  = \sum_{S_n/Z_n}   N_c \Tr[ T^{a_1} \cdots T^{a_n}] A_{n:1}^{(1)} (a_1,a_2,\cdots, a_n)
\notag
\\
&+&\sum_{r=2}^{[n/2]+1} \sum_{S_n/(Z_{r-1}\times Z_{n+1-r})}
\hskip-1.0truecm
\Tr[ T^{a_1} \cdots T^{a_{r-1}}]\Tr[T^{b_r} 
\cdots T^{b_{n}}] A_{n:r}^{(1)}(a_1,\cdots, a_{r-1} ; b_r, \cdots, b_{n})\,.
\label{eq:oneloopcolordeco}
\end{eqnarray}
The $A_{n:2}^{(1)}$ are absent (or zero) in the $SU(N_c)$ case.
For $n$ even and $r-1=n/2$ there is an extra $Z_2$ in the summation to ensure each color structure only appears once.
The partial amplitudes  $A_{n:r}^{(1)}(a_1,\cdots, a_{r-1} ; b_r, \cdots, b_{n})$ are cyclically symmetric in the sets $\{a_1,\cdots, a_{r-1} \}$ and 
$\{b_r, \cdots, b_{n} \}$ and obey a ``flip'' symmetry,
\begin{equation}
A^{(1)}_{n:r} (1,2,\cdots ,(r-1) ; r, \cdots ,n) =(-1)^n A^{(1)}_{n:r} (r-1,\cdots, 2,1 ; n, \cdots , r) \,.
\end{equation}  
Decoupling identities again impose relationships amongst the partial amplitudes.  For example setting  leg $1$ to be $U(1)$ and extracting the coefficient of 
$\Tr[T^2 T^3 \cdots T^n]$ implies
\begin{equation}
A_{n:2}^{(1)}(1;2,3,\cdots, n) +A_{n:1}^{(1)}(1,2,3,\cdots, n ) +A_{n:1}^{(1)}(2,1,3,\cdots, n )+\cdots A_{n:1}^{(1)}(2,\cdots, 1, n )=0
\label{eq:decoupleA}
  \end{equation}
and consequently $A_{n:2}^{(1)}$ can be expressed as a sum of $(n-1)$ of the $A_{n:1}^{(1)}$.  
By repeated application of the decoupling identities all the $A_{n:r}^{(1)}$ can be expressed as sums over the $A_{n:1}^{(1)}$~\cite{Bern:1990ux},
\begin{equation}
A_{n:r}^{(1)}(1,2,\ldots,r-1;r,r+1,\ldots,n)\ =\
 (-1)^{r-1} \sum_{\sigma\in OP\{\bar \alpha\}\{\beta\}} A_{n:1}^{(1)}(\sigma)
\label{eq_COP}
\end{equation}
where $\{\bar\alpha\} \equiv \{ r,r-1, \cdots , 1  \}$ and
$ \{\beta\} \equiv \{r,r+1,\ldots,n-1,n\}$. 
The set $OP\{S_1\}\{S_2\}$ is the set of all mergers of $S_1$ and $S_2$ which preserves the
order of $S_1$ and $S_2$ within the merged list. 
Consequently, at one loop only the leading order in color term need be computed. Unfortunately this feature does not persist beyond one-loop.

A general two-loop amplitude may be expanded in a color trace basis as
\begin{eqnarray}
& & {\cal A}_n^{(2)}(1,2,\cdots ,n) =
N_c^2 \sum_{S_n/Z_n}  \tr(T^{a_1}T^{a_2}\cdots T^{a_n}) A_{n:1}^{(2)}(a_1,a_2,\cdots ,a_n) \notag \\
&+&
N_c\sum_{r=2}^{[n/2]+1}\sum_{S_n/(Z_{r-1}\times Z_{n+1-r}) }   \tr(T^{a_1}T^{a_2}\cdots T^{a_{r-1}})\tr(T^{b_r} \cdots T^{b_n}) 
A_{n:r}^{(2)}(a_1,a_2,\cdots ,a_{r-1} ; b_{r}, \cdots, b_n)  
\notag \\
&+& \sum_{s=1}^{[n/3]} \sum_{t=s}^{[(n-s)/2]}\sum_{S_n/(Z_s\times Z_t \times Z_{n-s-t})} 
\hskip -1.0 truecm   \tr(T^{a_1}\cdots T^{a_s})\tr(T^{b_{s+1}} \cdots T^{b_{s+t}})
\tr(T^{c_{s+t+1}}\cdots T^{c_n}) 
\notag
\\
& & 
\hskip 7.0truecm 
\times A_{n:s,t}^{(2)}(a_1,\cdots ,a_s;b_{s+1}, \cdots, b_{s+t} ;c_{s+t+1},\cdots, c_n ) 
\notag \\
&+&\sum_{S_n/Z_n}  \tr(T^{a_1}T^{a_2}\cdots T^{a_n}) A_{n:1B }^{(2)}(a_1,a_2,\cdots ,a_n)\,.
\label{eq:twoloopexpansion}
\end{eqnarray}
Again, for $n$ even and $r-1=n/2$ there is an extra $Z_2$ in the summation to ensure each color structure only appears once.
In the $s,t$ summations there is an extra $Z_2$ when exactly two of $s$, $t$ and $n-s-t$ are equal and an extra $S_3$ when all three are equal. 

The focus of this article is the $A_{n:1B }^{(2)}$ term.
Decoupling identities do not relate the $A^{(2)}_{n:1B}$ to the other terms but do impose an identity analogous to that for the 
tree amplitude~eq.(\ref{eq:decotree}), 
\begin{equation}
A_{n:1B}^{(2)}(1,2,3,\cdots, n ) +A_{n:1B}^{(2)}(2,1,3,\cdots, n )+\cdots A_{n:1B}^{(2)}(2,\cdots, 1, n )=0\,.
\label{eq:decoupleB}\end{equation}
In itself this does not specify $A_{n:1B}^{(2)}$ completely. 
There are further relations amongst the $A^{(2)}_{n:\alpha}$ beyond the decoupling identities~\cite{Naculich:2011ep,Edison:2011ta} which may be obtained by recursive methods.  These relate $A^{(2)}_{n:\alpha}$ to other partial amplitudes and at five-point allow $A_{5:1B}^{(2)}$ to be
expressed in terms of the $A_{5:1}^{(2)}$ and $A_{5:3}^{(2)}$. However, beyond five point only $A_{6:1}^{(2)}$ and $A_{7:1}^{(2)}$ are currently known.

\section{A String Theory Interlude} 
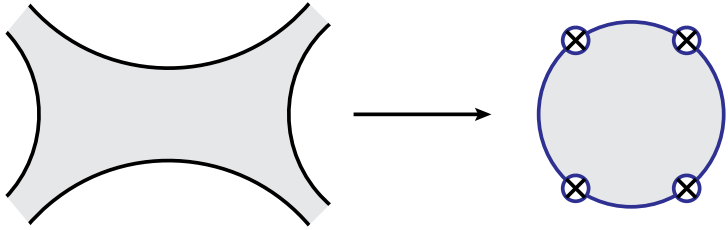
\begin{figure}[h]
\centerline{
\begin{picture}(300,150)(0,0)    
%\newcolor{LightGray}{0 0.0 0.0 0.05}
\SetScale{0.7}
\SetWidth{2}
\CBox(-50,10)(140,140){White}{LightGray}
\CCirc(50,200){100}{White}{White}
\CCirc(50,-50){100}{White}{White}
\CCirc(-85,75){65}{White}{White}
\CCirc(180,75){65}{White}{White}
\CArc(50,200)(100,220,320)
\CArc(50,-50)(100,40,140)
\CArc(-85,75)(65,-50,50)
\CArc(180,75)(65,130,230)
\LongArrow(150,75)(220,75) 
\COval(300,75)(50,50)(0){Blue}{LightGray}
\COval(270,115)(7,7)(0){Blue}{White}
\Line(265,120)(275,110)
\Line(265,110)(275,120)
\COval(330,115)(7,7)(0){Blue}{White}
\Line(325,120)(335,110)
\Line(325,110)(335,120)
\COval(330,35)(7,7)(0){Blue}{White}
\Line(325,30)(335,40)
\Line(325,40)(335,30)
\COval(270,35)(7,7)(0){Blue}{White}
\Line(265,30)(275,40)
\Line(265,40)(275,30)
%     \Line(60,60)(60, 0)
 %    \Line(60, 0)( 0, 0)
\SetColor{White}
\SetWidth{20}
\Line(-25,0)(-50,30)
\Line(-25,150)(-50,120)
\Line(125,0)(155,30)
\Line(125,150)(155,120)
  \end{picture} 
    }
    \caption{In open string theory, the surface linking external open string states may be mapped to a disc where the external states are vertex operators lying on the boundary. 
   }
    \label{fig:stringsurface}
\end{figure}

The partial amplitude $A^{(2)}_{n:1B}$ has an interesting source in open string theory.  
String theory contains massless gauge bosons as part of its spectrum 
of states and much can be gleaned from the string theory organisation of the scattering amplitudes.  An open string has endpoints 
with the quantum numbers of quarks and anti-quarks (Chan-Paton factors). The state thus lies in the adjoint of $U(N_c)$.   
A string amplitude is obtained by summing over all world sheets linking the external states. A simple example  is shown in  fig.~\ref{fig:stringsurface}.

The surface linking the external states can be conformally mapped to the surface shown with vertex operators attached to the
boundary. Each vertex operator contains an adjoint color matrix $T^a$.   Tracing over the color indices naturally gives an expansion of the amplitude in terms of color traces
\begin{equation}
A = \sum ( color traces   ) \times A(\alpha)
\end{equation}
where $\alpha$ is the string tension. The string theory amplitude contains contributions from an infinite number of states however in the infinite string tension limit the amplitude reduces to that of field theory. The colour structure survives this limit.

\begin{figure}[h]
\centerline{
\begin{picture}(200,100)(0,0)    
%\SetColor{Red}
\SetScale{0.7}
\SetWidth{2}
\newcolor{VLightGray}{0 0.0 0.0 0.02}
\COval(100,75)(70,120)(0){Red}{LightGray}
\COval(65,75)(30,30)(0){Green}{White}
\COval(135,75)(30,30)(0){Blue}{White}
%     \Line(60,60)(60, 0)
 %    \Line(60, 0)( 0, 0)
\SetColor{Black}
\SetWidth{1}
  \end{picture} 
    }
    \caption{A typical surface with three boundaries. Vertex operators can be attached to any of the boundaries.}
    \label{fig:twolooptypical}
\end{figure}
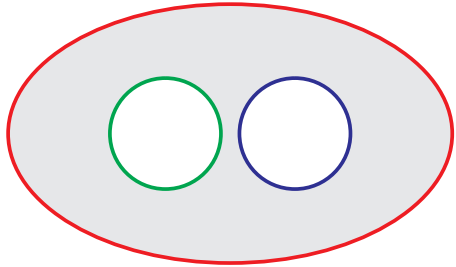
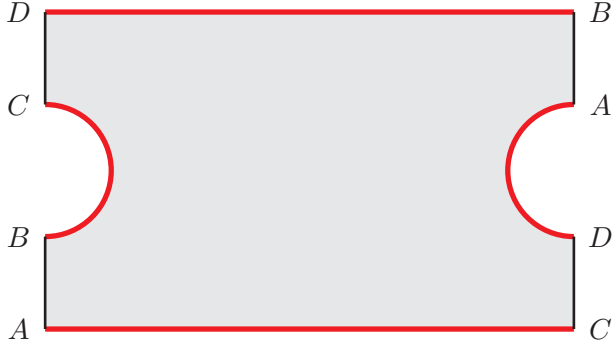
\begin{figure}[h]
\centerline{
\begin{picture}(200,150)(0,0)    
\SetScale{1.0}
\CBox(0,0)(200,120){White}{LightGray}
\COval(0,60)(25,25)(0){White}{White}
\COval(200,60)(25,25)(0){White}{White}
\SetColor{Red}
\SetWidth{2}
    \Line( 0, 0)( 200,0)
     \Line(0,120)(200,120)
 \CArc(0,60)(25,270,90)   
 \CArc(200,60)(25,90,270)    
%     \Line(60,60)(60, 0)
 %    \Line(60, 0)( 0, 0)
\SetColor{Black}
\SetWidth{1}
\Line(0,0)(0,35) 
\Line(0,120)(0,85)
\Line(200,0)(200,35) 
\Line(200,120)(200,85)
\Text(-10,0)[h]{$A$}
\Text(-10,35)[h]{$B$}
\Text(-10,85)[h]{$C$}
\Text(-10,120)[h]{$D$}
\Text(210,0)[h]{$C$}
\Text(210,35)[h]{$D$}
\Text(210,85)[h]{$A$}
\Text(210,120)[h]{$B$}   \end{picture} 
    }
    \caption{This surface with edges $A-B$ and $C-D$ identified is an oriented surface with a single edge. In string theory
    attaching vector bosons to the edge of this surface generates the sub-sub-leading single trace color term. }
    \label{fig:twoloopfig}
\end{figure}

A typical surface contributing at two-loop is shown in fig.~\ref{fig:twolooptypical}.  This has three boundaries to which
gauge boson  vertex operators may be attached. If no gauge bosons are attached a  factor of $N_c$ is generated by summing over the colors
the boundary may have. Populating this surface by vertex operators generates the expansion of eq.(\ref{eq:twoloopexpansion}) 
{\it except} for the single trace term $A^{(2)}_{n:1B}$.   This arises from a different category of surface.
If we consider the surface shown in fig.~\ref{fig:twoloopfig} with the edges identified as shown then the surface is a two-loop surface which is non-planar  but nonetheless is oriented and has a single boundary.  Attaching gauge bosons to the edge gives the single trace term and is,
in string theory, the source of $A^{(2)}_{n:1B}$.

%  } % HHHH
  
\section{The All-plus Amplitudes}

We are now in a position to look at the specific amplitude where all gluons have the same helicity.  
This particular amplitude vanishes at tree level:
\begin{equation}
A_{n}^{(0)}(1^+,2^+,\ldots,n^+)\ =\ 0  \; . 
\end{equation}
Consequently, the one-loop amplitude is rational (to order $\varepsilon^0$ in the dimensional regularisation parameter) and the two-loop amplitudes will have a simpler 
singular structure in $\varepsilon$.

The leading in color one-loop partial amplitude has an all-$n$ 
expression~\cite{Bern:1993qk}
\footnote{Here  a null momentum is represented as a
pair of two component spinors $p^\mu =\sigma^\mu_{\alpha\dot\alpha}
\lambda^{\alpha}\bar\lambda^{\dot\alpha}$. 
We are using a spinor helicity formalism with the usual
spinor products  $\spa{a}.{b}=\epsilon_{\alpha\beta}
\lambda_a^\alpha \lambda_b^{\beta}$  and 
 $\spb{a}.{b}=-\epsilon_{\dot\alpha\dot\beta} \bar\lambda_a^{\dot\alpha} \bar\lambda_b^{\dot\beta}$. 
\noindent{Also}
 $ s_{ab}=(k_a+k_b)^2=\spa{a}.b \spb{b}.a=\la a|b|a]$
 and
 $K_{ab\cdots r}=k_a+k_b\cdots +k_r$.
}

\begin{equation}
A_{n:1}^{(1)}(1^+,2^+,\ldots,n^+)\ =\ -{i \over 3}\,
{ 1\over \spa1.2 \spa2.3 \cdots \spa{n}.1} { \sum_{1\leq i < j < k < l  \leq n}{\rm tr}_-[i j k l ]  } 
+ O(\varepsilon) \,.
\label{eq:oneloopleading}
\end{equation}
This expression is order $\varepsilon^0$ but all-$\varepsilon$ expressions exist for the first few amplitudes in this series~\cite{Bern:1996ja}. 
In this expression, 
\begin{equation}
{\rm tr}_-[ijkl]\equiv {\rm tr}( \frac{(1-\gamma_5)}{2} \ksl_{i} \ksl_{j} \ksl_{k} \ksl_{l} ) 
=\frac{1}{2}{\rm tr}( \ \ksl_{i} \ksl_{j} \ksl_{k} \ksl_{l} ) -\frac{1}{2}\epsilon(i,j,k,l)
=\spa{i}.{j}\spb{j}.{k}\spa{k}.{l}\spb{l}.{i}
\end{equation}
and $\epsilon(i,j,k,l)={\rm tr}_+[ijkl]-{\rm tr}_-[ijkl]$.  
This amplitude has the same denominator as the Parke-Taylor amplitude. This combination will reappear in many places so we define
\begin{equation}
\PT (a_1,a_2,a_3,\cdots, a_n ) \equiv { 1\over \spa{a_1}.{a_2}\spa{a_2}.{a_3} \cdots \spa{a_n}.{a_1} }  
\equiv{1\over \CY( a_1,a_2,a_3,\cdots, a_n  )} \,.
\end{equation}
The numerator of eq.~(\ref{eq:oneloopleading}) can be split into trace terms and  $\epsilon$ pieces (originally called $E_n$ and $O_n$ in ref~\cite{Bern:1993qk}). 
Specifically for the five point amplitude,
\begin{equation}
A_{5:1}^{(1)}(1^+,2^+,3^+,4^+,5^+)\ =\ -{i \over 3}\,
{ { s_{12} s_{23}+s_{23} s_{34}+s_{34} s_{45}+s_{45} s_{51}+s_{51} s_{12} + \epsilon(1,2,3,4)   } 
\over \spa1.2 \spa2.3 \spa3.4\spa4.5\spa5.1 } 
+ O(\varepsilon) \,.
\end{equation}
The $\epsilon$ part of eq.(\ref{eq:oneloopleading}) will reappear later in a two loop amplitude.  The expression~eq.(\ref{eq:oneloopleading}) was first conjectured by 
studying  collinear limits starting with $n=5$ and later proven correct using off-shell recursion~\cite{Mahlon:1993si}.  

In~\cite{Dunbar:2019fcq}, we presented compact expressions for the subleading terms 
\begin{align}
A_{n:2}^{(1)}(1^+ ;2^+,3^+,\cdots, n^+) &= -i{ 1 \over \spa2.3\spa3.4 \cdots \spa{n}.2} \sum_{2 \leq i<j \leq n}  \spb{1}.{i} \spa{i}.{j} \spb{j}.{1} 
\notag \\
&= -i{\sum_{2 \leq i<j \leq n}  \spb{1}.{i} \spa{i}.{j} \spb{j}.{1} \over \CY(2,3,\cdots, n ) }
%\notag
\intertext{and for $r\ge3$}
A_{n:r}^{(1)}(1^+,2^+,\cdots, r-1^+ ;  r^+, \cdots, n^+) &= -2i {  (K_{1\cdots r-1}^2)^2   \over 
(\spa1.2\spa2.3 \cdots \spa{(r-1)}.1)  ( \spa{r}.{(r+1)}  \cdots \spa{n}.{r}  )} 
\notag \\
&= -2i { (K_{1\cdots r-1}^2)^2   \over \CY(1,2,\cdots ,r-1 ) \CY(r,r+1,\cdots ,n ) } \,.
\end{align}
These expressions are remarkably simple given the number of terms arising in the naive application of (\ref{eq_COP}).

At two loop, the all-plus amplitude has been computed for four and five points, its relative simplicity making it the first target in computations.  
At two loop the all-plus amplitude contains ``Infra-Red'' (IR) and ``Ultra-Violet'' (UV) infinities together with finite polylogarithmic and rational terms.
The IR singular structure of a color partial amplitude is determined by general theorems~\cite{Catani:1998bh}.  
Consequently we can split the amplitude into a term containing both the IR and UV divergences, $U^{(2)}_{n:\lambda}$,  and finite terms $F^{(2)}_{n:\lambda}$,
\begin{eqnarray}
\label{definitionremainder}
A^{(2)}_{n:\lambda} =& U^{(2)}_{n:\lambda}
+  \; F^{(2)}_{n:\lambda}  +   {\mathcal O}(\varepsilon)
\end{eqnarray}
($F^{(2)}_{n:\lambda}$ is the ``infrared finite hard'' function of ref.~\cite{Badger:2019djh}).

As the all-plus tree amplitude vanishes, $U^{(2)}_{n:\lambda}$ simplifies considerably and is only $1/\varepsilon^2$.
In general an amplitude has UV divergences, collinear IR divergences and soft IR divergences. 
As the tree amplitude vanishes,  both the UV divergences and collinear IR divergences are proportional to $n$ and cancel leaving only the 
soft IR singular terms~\cite{Kunszt:1994np}.

The leading IR singularity for the $n$-point two-loop amplitude is~\cite{Bern:2000dn}
\begin{equation}
-\frac{s_{ab}^{-\varepsilon}}{\varepsilon^2} f^{aij}f^{bik} \times {\cal A}_n^{(1)}(j,k,\cdots, n )
\end{equation}
where ${\cal A}_n^{(1)}$ is the full-color one-loop amplitude.  
We wish to disentangle this simple equation into the color-ordered partial amplitudes.  This was done for all two-loop 
colour amplitudes in ref.~\cite{Dunbar:2019fcq}: we reproduce the result for $A^{(2)}_{n:1B}$ here. 
Defining 
\begin{equation}
I_{i,j} \equiv -\frac{s_{ij}^{-\varepsilon}}{\varepsilon^2}
\end{equation}
and 
\begin{eqnarray}
I_k[ S_1,S_2]=I_k [ \{a_1,a_2\cdots a_r \}, \{b_1,b_2,\cdots b_s\} ]  \equiv
\left( I_{a_1,b_s}+I_{b_1,a_r} -I_{a_1,b_1} -I_{a_r,b_s}  \right)
\end{eqnarray}
then
\begin{align}
U_{n:1B}^{(2)}  (S) &= \sum_{Q(S)}  A_{n:r+1}^{(1)}  (S'_1;S'_2) 
\times  I_k[S'_1,S'_2]    \,, 
\label{eq:IRhard}
\end{align}
where $Q(S)$ is the set of all distinct pairs of lists satisfying $S'_1\oplus S'_2 \in C(S)$  where the size of $S'_i$ is greater than one 
and the set $C(S)$ is the set of cyclic permutations of $S$. 
For example
\begin{align}
Q(\{1,2,3,4,5\})= \biggl\{ &
(\{1,2\},\{3,4,5\} ),  
(\{2,3\},\{4,5,1\} ), 
(\{3,4\},\{5,1,2\} ), 
\notag \\ &
(\{4,5\},\{1,2,3\} ), 
(\{5,1\},\{2,3,4\} ) 
\biggr\}\,.
\end{align}
In eq.(\ref{eq:IRhard}), the $A_{n:r+1}^{(1)}$ are  the all-$\varepsilon$ forms of the one loop amplitude which can be specified by 
eq.~(\ref{eq_COP}).   These are only available in functional form for $n\leq 6$.

Given the general expressions for $U_{n:\lambda}^{(2)}$, the challenge is to compute the finite parts of the amplitude: $F_{n:\lambda}^{(2)}$. 
This finite remainder function $F_{n:\lambda}^{(2)}$ can be further split into polylogarithmic and rational pieces,
\begin{equation}
F_{n:\lambda}^{(2)} = \Fcc_{n:\lambda}+R_{n:\lambda}^{(2)}\; .
\end{equation}
We calculate the polylogarithmic piece using four-dimensional unitarity and the rational term using the factorisation properties of the amplitude
which we will discuss in the following section.

\section{Factorisation Properties of $A^{(2)}_{n:1B}$}

In this section we make some comments regarding the singularity structure of the sub-sub leading amplitudes: $A_{n:1B}^{(2)}$ and 
$A_{n:s,t}^{(2)}$.      
In general amplitudes  have

$a)$ Multiparticle Poles

$b)$ Double Complex Poles

$c)$ Complex Poles 

$d)$ Collinear Poles 

\noindent
We will demonstrate that  $A_{n:1B}$ is lacking the first two and that only the last is determined by general theorems. Fortunately this will be sufficient to generate a form for the rational functions. 

As the all-plus amplitude vanishes at tree level, multiparticle poles can only arise if the amplitude factorises into two one-loop factors,   
\begin{equation}
{\cal A}^{1-loop}  (\cdots , K^\lambda_i ) \times {1\over K^2 } \times {\cal A}^{1-loop} (\cdots, -K^{-\lambda}_i)\,.
\end{equation}
This is non-zero with one amplitude being the single minus one-loop amplitude and the other the all-plus. Both of these are rational. 
Only the subleading amplitudes from each of the one-loop factors will contribute to
the  $N_c^0$ term and the colour terms must be of the form
\begin{equation}
\sim \tr( i S_1 )\tr(S_2) \times \tr(i S_3)\tr(S_4)
\end{equation}
where we sum over the color matrix $T^i$ and we have suppressed the explicit colour matrices for the lists of legs $S_i$.   
The $S_1$ and $S_3$ may be null and  if both are null, we obtain a factor of $N_c$.  
Otherwise we obtain
\begin{equation}
\tr(S_1S_3)\tr(S_2)\tr(S_4)\,.
\end{equation}
So there are (1-loop)-(1-loop) factorisations in $A_{n:s,t}^{(2)}$ but not in $A_{n:1B}^{(2)}$.   
Therefore $A^{(2)}_{n:1B}$  has no $1/K^2$ terms.  The presence of the single minus amplitude within a limit would make it difficult to find an all-$n$ expression.

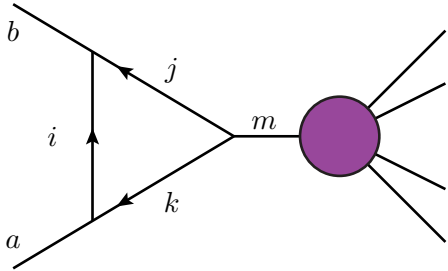
\begin{figure}[H]
\centerline{
         \begin{picture}(120,100)(0,0)    
     %\Line( 0, 0)( 0,60)
 \SetWidth{1.0}
     \ArrowLine(83.5,50)(0, 100)  
     \ArrowLine(83.5,50)(0, 0)
     \ArrowLine(30,18)(30,82) 
          \Line(83.3,50)(123,50)     
      \Line(123.5,50)(163.5,90)    
           \Line(123.5,50)(163.5,70)    
                \Line(123.5,50)(163.5,30)    
       \Line(123.5,50)(163.5,10)    
     \CCirc(123.5,50){15}{Black}{Purple} 
     \Text(60, 25)[c]{$k$}
     \Text(60, 75)[c]{$j$}
     \Text(15, 50)[c]{$i$}
          \Text(95, 55)[c]{$m$}
       \Text( 0,10)[c]{$a$}
         \Text( 0,90)[c]{$b$}           
     %\DashLine(25,50)(35,50){1}
     %\DashLine(55.1,63.5)(58.1,68.5){1}
     %\DashLine(55.1,36.5)(58.1,31.5){1}
                 \SetPFont{Arial}{5}
                 \SetColor{Red}
         \end{picture}    
     }
    \caption{Contributions to amplitudes giving a double pole with color indices shown.}
    \label{fig:double pole}
\end{figure}

Amplitudes also contain double poles in complex momentum. These arise from diagrams such as shown in fig.~\ref{fig:double pole} where one factor arises from the explicit pole and the other from the loop integral.
The color structure of the double pole diagram therefore contains
\begin{equation}
f^{a i k} f^{bij} f^{kjm}  J(m,\cdots ) \,.
\end{equation}
We can turn this into color traces and evaluate:
\begin{eqnarray*}
 & & \left(   \Tr[aki]-\Tr[kai] \right)\left( \Tr[bji]-\Tr[jbi] \right)\left( \Tr[kjm]-\Tr[kmj] \right)
\\
&=&   N_c\Tr [bam]-N_c\Tr[abm] \,.
\end{eqnarray*}
Hence there is no $N_c^0$ contribution and $A^{(2)}_{n:1B}$ is free of double poles.

Unfortunately, the single poles are not as simple as one might imagine.  For example, at five point the potential factorisation
\begin{equation}
A_{5:1B}^{(2)} \longrightarrow  A_{3}^{(0}(a^+,b^+,K^-)  \times \frac{1}{s_{ab} } \times A_4^{(2)} (K^+, \cdots )
\end{equation}
vanishes since $A_4^{(2)} (1^+,2^+,3^+,4^+)=0$, nonetheless $A_{5:1B}^{(2)}$ in eq.~(\ref{eq:R5B}) has poles in
$\spa{a}.{b}$. 
These single poles arise from non-factorisating terms as computed 
in~\cite{Dunbar:2019fcq,Dalgleish:2020mof} where the double and single poles are  determined for the $n=5$ and $n=6$ cases.  

Finally let us consider collinear limits.  If adjacent legs $a$ and $b$ become collinear with $k_a=zK$ and $k_b=(1-z)K$, then we expect
\begin{equation}
A_{n:1B}^{(2)} ( \cdots , a^+ , b^+ , \cdots ) \longrightarrow   S^{++}_{-} ( a,b, K) A_{n-1:1B}^{(2)} ( \cdots ,  K^+ , \cdots )
\end{equation}
where
\begin{equation}
S^{++}_{-} ( a,b, K)  = {1\over \sqrt{z(1-z)} \spa{a}.b } \; . 
\end{equation}
The amplitude has no collinear singularity  if $a$ and $b$ are not adjacent. 
Demanding the correct collinear behaviour was sufficient to generate the conjecture for the one-loop all plus amplitude.

\section{Polylogarithic Terms }

In refs.~\citep{Dunbar:2016aux,Dunbar:2016cxp,Dunbar:2016gjb,Dunbar:2017nfy}  it was demonstrated that for the leading in color partial amplitude the 
IR infinite terms and the polylogarithmic terms may  be generated using 
four dimensional unitarity cuts~\cite{Bern:1994zx,Bern:1994cg}. In particular quadruple cuts~\cite{Britto:2004nc} could be used to compute the 
coefficients of functions which were essentially the  finite parts of one loop box functions.  

 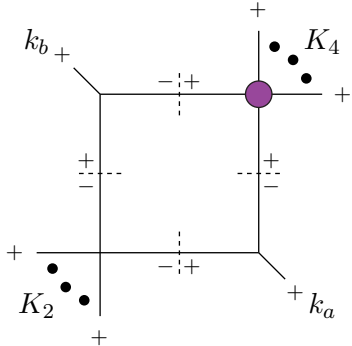
\begin{figure}[H]
\centerline{
    \begin{picture}(170,150)(-50,-50) 
     \Line( 0, -24)( 0,60)
     \Text(0,-32)[c]{$_+$}
     \Text(-14,75)[c]{$_+$}
     \Text(74,-15)[c]{$_+$}
     \Text(-32,0)[c]{$_+$}
\Text(-24,-20)[c]{$K_2$}
\Text(84,-20)[c]{$k_a$}
\Text(-24,80)[c]{$k_b$}
\Text(84,80)[c]{$K_4$}
\Vertex(-6,-18){2}
     \Vertex(-18,-6){2}
     \Vertex(-13,-13){2}
     \Line( 0,60)(60,60)
     \Line(60,60)(60, 0)
     \Line(60, 0)( -24, 0)
     \Line( 0,60)(-10,70)
     \Line(60,60)(60,84)
     \Line(60,60)(84,60)
     \Text(60,92)[c]{$_+$}
     \Text(92,60)[c]{$_+$}
     \Vertex(66,78){2}
     \Vertex(78,66){2}
     \Vertex(73,73){2}
     \Line(60, 0)(70,-10)
     \DashLine(52,30)(68,30){2}
     \DashLine(-8,30)(8,30){2}
     \DashLine(30,8)(30,-8){2}
     \DashLine(30,52)(30,68){2}
      \Text(-5,25)[c]{$_-$}
      \Text(-5,35)[c]{$_+$}
      \Text(25,-5)[c]{$_-$}
      \Text(35,-5)[c]{$_+$}
      \Text(65,25)[c]{$_-$}
      \Text(65,35)[c]{$_+$}
      \Text(25,65)[c]{$_-$}
      \Text(35,65)[c]{$_+$}
   \CCirc(60,60){5}{Black}{Purple} 
    \end{picture}  
    }
    \caption{Four dimensional cuts of the two-loop all-plus amplitude involving an all-plus one-loop vertex 
    (indicated by $\bullet\;$ ). $K_2$ may be null but $K_4$ must contain at least two external legs. } 
    \label{fig:oneloopstyle}
\end{figure}

\def\plF{{\rm F}}

The expression for the $\Fcc_{n:\lambda}$ for the all-plus color amplitudes is~\cite{Dunbar:2016cxp} of the form
\begin{equation}
P_{n:\lambda}^{(2)}   =  \sum_{i}   c_{i}  \plF^{2m}_{i}
\end{equation}
where $c_i$ are rational functions and 
\begin{eqnarray}
\plF^{\rm 2m }[S,T,K^2_2,K_4^2]  &=& 
% \notag \\ 
  \Li_2\left(1-{ K_2^2 \over S }\right)
   +\ \Li_2\left(1-{K_2^2 \over T}\right)
 + \Li_2\left(1-{ K_4^2 \over S }\right)
   + \ \Li_2\left(1-{K_4^2 \over T}\right)  
\notag \\ 
& &  -\Li_2\left(1-{K_2^2K_4^2\over S T}\right)
   +{1\over 2} \ln^2\left({ S  \over T}\right)
   \; .
\end{eqnarray}
The $\plF^{\rm 2m}$ are the combination of polylogs which appear in the two-mass box with the orientation of fig.~\ref{fig:oneloopstyle}
with $S=(K_2+k_a)^2$ and $T=(K_2+k_b)^2$. 
In the specific case where $K_2^2=0$, 
\begin{eqnarray}
 \plF^{\rm 2m }[S,T,0,K_4^2] &=& 
% \notag \\ 
 \Li_2\left(1-{ K_4^2 \over S }\right)
   + \ \Li_2\left(1-{K_4^2 \over T}\right)  
   +{1\over 2} \ln^2\left({ S  \over T}\right)+{\pi^2 \over 6} \; .
\end{eqnarray}

For $A_{n:1B}^{(2)}(1^+,\cdots, n^+)$ we will need specific combinations which we label
\begin{equation}
\plF(a,b; S_2 ; S_4)=\plF^{\rm 2m }[K^2_{aS_2},K^2_{S_2b},K^2_{S_2},K^2_{S_4}]
\end{equation}
where $S_2$ are the set of external legs within $K_2$ and $S_4$ are the set of legs within $K_4$.  
With this we have 
\begin{eqnarray}
P_{n:1B}^{(2)} =-2i \sum_{a < b}   \Biggl( & & 
\sum_{(U_1^i:U_2^i )\in Spl_2(U_{ab})} 
\sum_{(V_1^j:V_2^j )\in Spl_2(V_{ab})}  c({a,b,U_1^i,V_1^j,U_2^i,V_2^j} ) \plF(a,b ; U_1^i\cup V_1^j; U_2^i \cup V_2^j)
\notag\\ & & 
+\sum_{(U_1^i:U_2^i )\in Spl_2(U_{ab})} 
\sum_{(V_1^j:V_2^j )\in Spl_2(V_{ab})}  c({a,b,U_2^i,V_2^j,V_1^j,U_1^i} ) \plF(a,b ; V_2^j \cup U_2^i ; U_1^i \cup V_1^j)
\notag\\ & & 
-\sum_{(V_1^i:V_2^i:V_3^i )\in Spl_3(V_{ab})} c({a,b,U_{ab},V_2^i,V_1^i,V_3^i} ) \plF(a,b ; U_{ab}\cup  V_2^i ; V_1^i \cup V_3^i)
\notag\\ & & 
-\sum_{(U_1^i:U_2^i:U_3^i )\in Spl_3(U_{ab})} c({a,b,U_2^i,V_{ab},U_3^i,U_1^i} ) \plF(a,b ; U_2^i \cup V_{ab} ; U_3^i \cup U_1^i)
\Biggl)
\end{eqnarray}
\def\PT{C_{PT}}
where
\begin{equation}
c({a,b,A_1,A_2,B_1,B_2})\equiv \la a |K_{B_2}K_{B_1}|b\ra^2 \PT ( a A_1 b A_2 )\PT ( b B_1) \PT( B_2 a)
\end{equation}
provided the $B_i$  and  $A_1\cup A_2$ are not null
\footnote{
For clarity we have suppressed list notation, so that $\PT ( b B_1)=\PT ( b, {B_1}_1 ,{B_1}_2, \cdots,{B_1}_r)$ etc.
}
%. $c({a,b,A_1,A_2,B_1,B_2})$ is defined to be zero otherwise. 
Also,
\begin{equation}
U_{ab}= \{ a+1,a+2,\cdots , b-1 \} \; {\rm and} \;  V_{ab}=\{ b+1, b+2, \cdots, n, 1, \cdots, a-1 \}
\end{equation}
i.e. the list $\{1,2,\cdots n \}$ is split, after cycling to begin with $a$, 
\begin{equation}
\{1,2,\cdots, n \}  \longrightarrow  \{ a, U, b,  V \}  \; .
\end{equation}
$Spl_2$ is the set of splits of a list  into two lists maintaining list order . So if 
$U=\{u_1, u_2, \cdots u_r\}$ then 
\begin{equation}
Spl_2 (U)=\{ U^i\} , U^i = ( \{u_1, \cdots u_i \} ; \{u_{i+1}, \cdots, u_r \} )
\end{equation}
and similarly $Spl_3(U)$ is the set of three lists obtained by splitting $U$ into three lists whilst maintaining order. 

The above expression is quite complex but simplifies significantly for small numbers of legs.  The sets $U_{ab}$ and $V_{ab}$ get split into two or three subsets which then get recombined into the sets of legs forming $K_2$ and $K_4$. For small $n$ many of the summations become trivial. 

In ref.~\cite{Henn:2019mvc} the one-loop all-plus amplitude was shown to be conformally invariant. In doing so the one-loop amplitude was rewritten to make the conformal symmetry manifest by writing the amplitude~(\ref{eq:oneloopleading}) as a sum of "$C_{kmn}$'' terms each of which is individually conformally invariant.
Writing the amplitude in terms of the $C_{kmn}$ terms occurs in a string theory based approach~\cite{Mafra:2012kh,He:2015wgf}.

The coefficients $c({a,b,A_1,A_2,B_1,B_2})$ are similar in structure although not identical to the $C_{kmn}$. They are however also
conformally invariant. We have verified that the conformal operator $k_{\alpha\dot\alpha}$ annihilates these. 
Specifically,\footnote{The $\lambda_{\alpha}^i$ and $\bar\lambda_{\dot\alpha}^i$ are not independent variables but satisfy 
$\sum_i \lambda_{\alpha}^i \bar\lambda_{\dot\alpha}^i=0$. We can either eliminate the dependant variables before applying the $k_{\alpha\dot\alpha}$ operator or include a $\delta(\sum_i \lambda_{\alpha}^i \bar\lambda_{\dot\alpha}^i)$ function. We have chosen the former route and checked eq.~(\ref{eq:conf}) at explicit kinematic points.} 
\begin{equation}
k_{\alpha\dot\alpha} c({a,b,A_1,A_2,B_1,B_2}) \equiv 
\left(  \sum_{i=1}^n { \partial^2 \over \partial \lambda_{\alpha}^i \partial \bar\lambda_{\dot \alpha}^i } 
\right)  c({a,b,A_1,A_2,B_1,B_2})
=0 \,.
\label{eq:conf}
\end{equation} 
The conformal invariance  of the coefficient of the polylogarithms was noted for the five-point amplitude in ref.~\cite{Henn:2019mvc}.

\section{Explicit Formula of $R^{(2)}_{n:1B}$}

The four point amplitude $R_{4:1B}^{(2)}$ has been calculated in~\cite{Glover:2001af,Bern:2002tk}
as part of the full four point computation and found to vanish: 
\begin{equation}
   R^{(2)}_{4:1B}(1^+,2^+,3^+,4^+)= 0\,.
\end{equation}

The five point amplitude has been computed.   
In~\cite{Badger:2019djh} ,  five point amplitudes $A^{(2)}_{5:1}$ and $A^{(2)}_{5:3}$ were computed explicitly.  Using the 
results of~\cite{Edison:2011ta} 
this implies a form of $A_{5:1B}^{(2)}$.   In~\cite{Dunbar:2019fcq} the $A^{(2)}_{5:r}$ were recomputed using augmented recursion~\cite{Dunbar:2010xk,Dunbar:2017nfy} 
and four dimensional unitarity and $A_{5:1B}^{(2)}$ was computed directly in a simple form.  
The explicit form is
\begin{align}
    R^{(2)}_{5:1B}(1^+,2^+,3^+,4^+,5^+)&=2i\epsilon\left(1,2,3,4\right) \sum_{Z_5(1,2,3,4,5)}
    \text{C}_{\text{PT}}(1,2,5,3,4)  
\notag \\    =2i\epsilon\left(1,2,3,4\right) 
\Biggl(
  &  
  \text{C}_{\text{PT}}(1,2,5,3,4)  +     
  \text{C}_{\text{PT}}(2,3,1,4,5)  +       
  \text{C}_{\text{PT}}(3,4,2,5,1)  
\notag  \\
 &  \hskip 3.0 truecm +         
   \text{C}_{\text{PT}}(4,5,3,1,2)  +
   \text{C}_{\text{PT}}(5,1,4,2,3)  
                \Biggr)
\label{eq:R5B}\end{align}
Since the summation is over the five cyclic permutations of the legs $(1,2,3,4,5)$ this expression is manifestly cyclically symmetric. 
However it is far from unique since the Parke-Taylor factors  $\text{C}_{\text{PT}}$ are not all linearly independent.   Since they are manifestly 
cyclic symmetric  there
are clearly $(n-1)!$ terms. They also satisfy identities identical to the decoupling identity for tree amplitudes which can be 
used to reduce these to a basis of $(n-2)!$ independent terms. Specifically we can rewrite
\begin{equation}
\sum_{(a_2,a_3,\cdot, a_n)\in P(2,3,\cdots, n)}  \alpha_i \text{C}_{\text{PT}} (1, a_2,a_3,\cdots, a_n)  
=\sum_{(a_2,a_3,\cdot, a_{n-1})\in P(2,3,\cdots, {n-1})}\alpha'_i \text{C}_{\text{PT}} (1, a_2,a_3,\cdots ,a_{n-1}, n)  
\end{equation}
If we choose to rewrite $R^{(2)}_{n:1B}$ in terms of this reduced set, cyclic symmetry will not be manifest but there is the advantage of working 
with a basis rather than a spanning set.   For the five point amplitude we then have
\begin{eqnarray}
   R^{(2)}_{5:1B}(1^+,2^+,3^+,4^+,5^+) & &=2i\epsilon\left(1,2,3,4\right)  \Biggl(  -\text{C}_{\text{PT}} (1,2,3,4,5)  
\\
& & +2\Bigl(\text{C}_{\text{PT}} (1,3,4,2,5)  +\text{C}_{\text{PT}} (1,4,3,2,5)  +\text{C}_{\text{PT}} (1,4,2,3,5)  
\Bigr)\Biggr)
\notag 
\end{eqnarray}
This can be split into two parts
\begin{eqnarray}
   R^{(2)}_{5:1B}(1^+,2^+,3^+,4^+,5^+) & &
  =  R^{(2)}_{5:1B_1}(1^+,2^+,3^+,4^+,5^+)  +   R^{(2)}_{5:1B_2}(1^+,2^+,3^+,4^+,5^+)
\end{eqnarray}
where
\begin{eqnarray}
   R^{(2)}_{5:1B_1}(1^+,2^+,3^+,4^+,5^+) & =&-2i\epsilon\left(1,2,3,4\right)  \text{C}_{\text{PT}} (1,2,3,4,5)  
\\
 R^{(2)}_{5:1B_2}(1^+,2^+,3^+,4^+,5^+) &=&
4i\epsilon\left(1,2,3,4\right) 
\left(\text{C}_{\text{PT}} (1,3,4,2,5)  +\text{C}_{\text{PT}} (1,4,3,2,5)  +\text{C}_{\text{PT}} (1,4,2,3,5)  
\right)
\notag
\end{eqnarray}

The term $R^{(2)}_{5:1B_1}$ is reminiscent of the one loop expression which allows us to propose the $n$-point expression
\begin{equation}
 R^{(2)}_{n:1B_1}(1^+,2^+,\cdots ,n^+) =-2i \, \text{C}_{\text{PT}} (1,2,\cdots ,n-1,n) \times\sum_{1\leq i < j < k < l \leq n } \epsilon(i,j,k,l)
\end{equation}
The expression for $R^{(2)}_{6:1B_2}$ has fourteen terms, 
\begin{eqnarray}
  & &R^{(2)}_{6:1B_2}(1^+,2^+,3^+,4^+,5^+,6^+)  =4i
 \Bigl(
{  \epsilon(3, 4, 5, 6) \over \CY(1, 2, 4, 5, 3, 6)  } + 
{   \epsilon(3, 4, 5, 6)\over  \CY(1, 2, 5, 3, 4, 6) }+ 
 {  \epsilon(3, 4, 5, 6)\over  \CY(1, 2, 5, 4, 3, 6)  }  
 \notag\\
 & &+ { \epsilon(1, 2, 3, 4)\over  \CY(1, 3, 4, 2, 5, 6)  } 
 -{ \epsilon(1, 2, 3, 6)\over  \CY(1, 3, 4, 5, 2, 6)  } + 
  { \epsilon(1, 2, 3, 4)\over  \CY(1, 4, 2, 3, 5, 6)  }  
 -{ \epsilon(1, 3, 4, 6)\over \CY(1, 4, 2, 5, 3, 6)  }  
 \notag \\
 & &  +{ \epsilon(1, 2, 3, 4) \over \CY(1, 4, 3, 2, 5, 6)  } + 
  { \epsilon(1, 2, 4, 6)\over \CY(1, 4, 3, 5, 2, 6)  }  
 -{ \epsilon(1, 3, 4, 6)\over \CY(1, 4, 5, 2, 3, 6)  } + 
  { \epsilon(1, 2, 4, 6)\over \CY(1, 4, 5, 3, 2, 6)  }  
\notag\\
 & & -{ \epsilon(1, 4, 5, 6)\over \CY(1, 5, 2, 3, 4, 6)  } + 
  { \epsilon(1, 3, 5, 6)\over \CY(1, 5, 2, 4, 3, 6)  } + 
  { \epsilon(1, 3, 5, 6)\over \CY(1, 5, 4, 2, 3, 6)  } 
 -{ \epsilon(1, 2, 5, 6)\over \CY(1, 5, 4, 3, 2, 6) }
\Bigr)
\; . 
\end{eqnarray}
This expression was first constructed by demanding it satisfy the correct collinear limits and subsequently verified using
augmented recursion techniques~\cite{Dalgleish:2020mof}.  

While this is the minimal expression, it is not the best for generalising. 
Defining
\begin{equation}
\epsilon(\{a_1,a_2,\cdots, a_m \}, b , c , \{d_1,d_2,\cdots, d_p\})
\equiv\sum_{i=1}^{m}\sum_{j=1}^p \epsilon(a_i , b , c , d_j)\,,
\end{equation}
 we can replace 
$\epsilon(3,4,5,6)$  by
$\epsilon(\{1,2\},4,3,6)$ etc. which makes the pattern clearer. 

Then by demanding the correct collinear limits we are led to the expression
\begin{eqnarray}
 & & R^{(2)}_{n:1B_2}(1^+,2^+,\cdots ,n^+) =4i \sum_{r=1}^{n-4}\sum_{s=r+4}^n
\notag \\
& & 
 \sum_{i=r+1}^{s-2}\sum_{j=i+1}^{s-1}
\epsilon(\{1,\cdots, r\},j,i,\{s,\cdots , n\} ) (-1)^{i-j+1}
\times \sum_{\alpha\in S_{r,s,i,j}}  \text{C}_{\text{PT}} (\{\alpha_{S_{r,s,i,j}} \} ) \,.
\end{eqnarray}

To define $S_{r,s,i,j}$ we divide the list of indices, 
\begin{align}
\{ 1,2,3,\cdots , n\} &=  \{ 1, \cdots ,r; r+1, \cdots, i-1; i ; i+1, \cdots , j-1; j; j+1,\cdots ,s-1; s,\cdots, n \}
\notag \\
&\equiv\{ 1, \cdots r, \} \oplus S_1 \oplus\{ i \} \oplus S_2  \oplus\{ j \} \oplus  S_3 \oplus \{ s, \cdots, n \}
\end{align}
with
\begin{equation}
S_1=\{r+1, \cdots ,i-1\}, \;\;\; S_2=\{  i+1, \cdots ,j-1\} , \;\;\ S_3=\{ j+1,\cdots ,s-1 \} \,.
\end{equation}
The sets $S_i$ may be null. Then 
\begin{equation}
S_{r,s,i,j} =  Mer( S_1, \bar S_2, S_3) 
\end{equation}
where $\bar S_2$ is the reverse of $S_2$ and $Mer( S_1, \bar S_2, S_3)$ is the set of all mergers of the three sets which respect the ordering
within the $S_i$ 
and
\begin{equation}
\alpha_{S_{r,s,i,j}} = \{ 1, \cdots, r \} \oplus \{ j \} \oplus \alpha \oplus \{ i \} \oplus \{ s, \cdots, n \}
\; . 
\end{equation}

The expression for $R^{(2)}_{n:1B_2}$ presumably has other realisations,  however within the chosen basis the 
coefficients of the $\text{C}_{\text{PT}}$  are uniquely given.   The expression has the correct collinear limit of legs $n-1$ and $n$
but does not have manifest cyclic symmetry however we have checked to a large number of external legs (up to 14) that the expression is  
cyclically symmetric, that it has all the correct collinear limits and  it has the correct flip properties.    The
$R^{(2)}_{n:1B_1}$ and $R^{(2)}_{n:1B_2}$ do not
individually satisfy the decoupling identity however the combination $R^{(2)}_{n:1B_1}+R^{(2)}_{n:1B_2}$ does. 

The term $R^{(2)}_{n:1B_1}$ can be rewritten in a form which looks more similar to $R^{(2)}_{n:1B_2}$ by manipulating the tensors 
\begin{eqnarray}
 & & R^{(2)}_{n:1B_1}( 1^+ , 2^+,\cdots ,n^+) 
 =
 -2i  \, \text{C}_{\text{PT}} (1,2,\cdots ,n)\times\sum_{1\leq i < j < k < l \leq n } \epsilon(i,j,k,l)
\\
&= &
-2i \, \text{C}_{\text{PT}} (1,2,\cdots ,n)\times\sum_{r=1}^{n-4}\sum_{s=r+4}^{n} \epsilon(\{1,2,\cdots, r\}, r+1,s-1,\{s,s+1,\cdots, n\}) 
 \; . 
 \notag
\end{eqnarray}
Although the coefficients of the polylogarithms are annihilated by the conformal operator $k_{\alpha\dot\alpha}$ we can confirm
\begin{equation}
k_{\alpha\dot\alpha} \left( R^{(2)}_{n:1B_1}( 1^+ , 2^+,\cdots ,n^+) +R^{(2)}_{n:1B_2}( 1^+ , 2^+,\cdots ,n^+) 
\right) \neq 0 \,.
\end{equation}

\section{Conclusions}

We have presented an ansatz for a very specific color amplitude at two loops which is valid for an arbitrary number of external legs.  Although we are short of a proof of the ansatz it satisfies consistency conditions and  factorisations which suggest it is correct.  
All-$n$ formulae provide a very
useful laboratory for testing conjectures and behaviour. For example, it was recently shown in ref.~\cite{Henn:2019mvc} that the one-loop all-plus
amplitude is conformally invariant: however the all-$n$ expression  allows us to check that $R_{n:1B}^{(2)}$ is {\it not} conformally invariant although the coefficients of the polylogarithms are. 
The all-plus amplitude 
at one-loop is very special and has relations to amplitudes in other theories. In particular the $N=4$ MHV amplitude is related to it by a dimension shift of integral functions~\cite{Bern:1996ja} and also the one-loop amplitude coincides with that of 
self-dual Yang-Mills~\cite{Cangemi:1996rx,Chalmers:1996rq}.   It would be very interesting to see if any of these or similar properties extend
to two-loop and beyond.

\section{Acknowledgements}
DCD was supported by STFC grant ST/L000369/1. JMWS was 
supported by STFC grant ST/S505778/1.

\appendix

%\vfill\eject

%\newpage

\bibliography{TwoLoop}{}

\bibliographystyle{h-physrev}
\end{document}